\documentclass[slac_one]{revtex4}
\usepackage{graphicx}
\usepackage{graphpap}
\usepackage{fancyhdr}
\begin{document}

\title{Neutralino annihilation to quarks with SUSY-QCD corrections} 

\author{K.~Kova\v{r}\'{i}k, B.~Herrmann}
\affiliation{Laboratoire de Physique Subatomique et de Cosmologie,\\ Universit\'{e} Joseph Fourier Grenoble, CNRS/IN2P3, Institut Polytechnique de Grenoble\\ 53 rue des Martyrs, 38026 Grenoble, France}

\begin{abstract}
	The calculation of the cosmological relic density of the dark matter
	candidate within supersymmetric models is an interesting possibility to
	obtain additional constraints on the supersymmetric parameter space with
	respect to collider, electroweak precision, and low-energy data. 	
	Considering future cosmological precision measurements, radiative corrections can play an important role in the analysis.
	We present full QCD and SUSY-QCD corrections to neutralino pair annihilation
	into quark-antiquark pairs and analyze their impact on the neutralino annihilation cross section and the relic density. 
	By comparing to the relic density obtained from a pure leading order
	calculation, we show that the corrections strongly influence the
	extraction of SUSY mass parameters from cosmological data.
\end{abstract}

\maketitle

\thispagestyle{fancy}


\section{INTRODUCTION} 
Beginning with the rotational curves of galaxies, the astronomical and astrophysical evidence gathered in recent years  supports the hypothesis of dark matter in our Universe. Moreover, the models of structure formation favor the so-called cold dark matter which consists of weakly interacting particles with non-relativistic velocities.
Based mainly on the cosmic microwave background measurements, one obtains the following limits on the relic density of cold dark matter  cross-section
\begin{equation}\label{cWMAP}
  0.095 < \Omega_{\rm CDM}h^2 < 0.136\,,
\end{equation}
where the $2\sigma$ interval was obtained from the three-year data of the Wilkinson Microwave Anisotropy Probe (WMAP), combined with the results of Sloan Digital Sky Survey (SDSS), Supernovae Legacy Survey (SNLS) experiment and from baryonic acoustic oscillations data \cite{WMAP}.
\newline %
One of the main motivations for physics beyond the Standard Model (SM) is the inability of SM to provide candidates for cold dark matter. In the Minimal Supersymmetric Standard Model (MSSM) with R-parity conservation, the lightest supersymmetric particle (LSP) is always stable. If it is also a colour singlet, electrically neutral and massive, it is a good candidate for cold dark matter. In many scenarios, in particular those where supersymmetry breaking is gravity mediated, the lightest neutralino is the LSP and a suitable dark matter candidate. We can then calculate its number density $n$ by solving the Boltzmann equation
\begin{equation}
  \frac{dn}{dt} ~=~ -3 H n - \langle \sigma_{\rm ann}v \rangle \big( n^2 - n_{\rm eq}^2 \big) ,
\end{equation}
where the first term on the right-hand side corresponds to a dilution due to the expansion of the Universe, and the second one to a decrease due to annihilations and co-annihilations. Here, $H$ denotes the Hubble expansion parameter, $n_{\rm eq}$ the density of the relic particle in thermal equilibrium and $v$ is the relative velocity of the annihilating pair. 
\newline %
The Boltzmann equation can be integrated numerically and the energy density of the dark matter particle is approximately inversely proportional to the thermally averaged annihilation cross-section $\langle \sigma_{\rm ann}v \rangle$,
\begin{equation}
  \Omega_{\rm CDM}h^2 ~\propto~ \frac{1}{\langle \sigma_{\rm ann}v \rangle}\,.
\end{equation}
The latter includes all annihilation and co-annihilation processes of the dark matter particle into Standard Model particles.
These by now standard calculations are incorporated in program packages such as {\tt DarkSUSY} \cite{Dsusy} and {\tt MicrOMEGAs} \cite{microm} which allow different scenarios and different models to be studied. 
\newline %
Using the experimental limits (\ref{cWMAP}), we can put constraints on the MSSM parameter space. If one is confined to the MSSM with gravity mediated supersymmetry breaking (mSUGRA), the constraints allow for a rather narrow allowed band in the space of the high-scale parameters $m_0, m_{1/2}, A_0, \tan\beta,{\rm sgn }\,\mu$. Moreover, the constraints are expected to become even more strict in the future when the Planck mission will provide additional data.  
\newline %
To match and fully exploit the current and expected experimental precision, one has to improve the theoretical prediction of the relic density by including the radiative corrections. We calculated the QCD and SUSY-QCD corrections to neutralino annihilation into quarks. In the work presented here, we investigate the effects of radiative corrections on the constraints of the MSSM parameter space.
\section{NEUTRALINO ANNIHILATION INTO QUARKS}
Neutralino annihilation into fermions is one of the most important annihilation channels. It is dominant for light neutralinos where all other final states are not kinematically allowed. The leading order cross-section was discussed in much detail in \cite{drees}. The cross-section in the non-relativistic limit is conventionally written as (see \cite{jung_rev})
\begin{equation}
v\sigma = a + bv^2 + \mathcal{O}(v^4)\,,	
\end{equation}
where $v$ is the relative velocity of the neutralino pair. In the case of annihilation into fermions, the coefficient $a$ is
proportional to the mass of the fermion \cite{drees}. As a consequence, the dominant final states
are the third generation fermions, i.e. in our case the top and bottom quarks. Note that the contribution from the coefficient $b$ is suppressed with respect to the one coming from $a$.
\newline %
The annihilation into fermions has to be enhanced by a resonance in order to yield a small enough relic density satisfying  the experimental constraints (\ref{cWMAP}). It proceeds mainly through a Higgs boson resonance (in particular the CP-odd Higgs resonance). This implies that Yukawa coupling of fermions to the Higgs boson becomes an essential parameter in this calculation along with $\tan\beta$. The bottom Yukawa coupling receives large SUSY corrections for large $\tan\beta$ \cite{carena}, which is another reason to incorporate radiative corrections in the predictions. The approximation used in \cite{carena} is already included in the {\tt MicrOMEGAs} package alongside with the QCD corrections for Higgs decays into fermions (see \cite{drehik}).
\begin{table}[t]
  \begin{center}
    \begin{tabular}{c||ccccc|c|cc||}
       & $m_0$ (GeV) & $m_{1/2}$ (GeV) & $A_0$ (GeV) & $\tan\beta$ & sgn($\mu$) & $\Omega_{\rm CDM}h^2$ & $b\bar{b}$ & $t\bar{t}$ \\
       \hline \hline
       1 & 1500 & 130 & -1500 & 10 & + & 0.116 & 86\% & -- \\
       2 & 1500 & 1500 & 0 & 50 & + & 0.112 & 83\% & --  \\
       \hline
       3 & 5300 & 625 & -1500 & 10 & + & 0.110 & -- & 72\% \\
       4 & 3000 & 600 & 0 & 50 & + & 0.110 & 15\% & 64\% \\
       \hline \hline
    \end{tabular}
  \end{center}
  \vspace*{-4mm}
  \caption{The minimal supergravity (mSUGRA) scenarios that lead to a neutralino relic density agreeing with the current limits of Eq.\ (\ref{cWMAP}), satisfy the current SUSY-particle mass limits, and present important contributions of neutralino annihilation into bottom or top quark-antiquark pairs. We indicate the mSUGRA parameters at the high scale, the relic density obtained with {\tt micrOMEGAs}, and the contributions of the different quark final states to the annihilation cross-section $\langle \sigma_{\rm ann}v \rangle$.}
\label{Tpoints}
\end{table}
\begin{figure}[t]
	\begin{picture}(450,290)(0,0)
		\put(5,150){\resizebox{!}{5.1cm}{\includegraphics{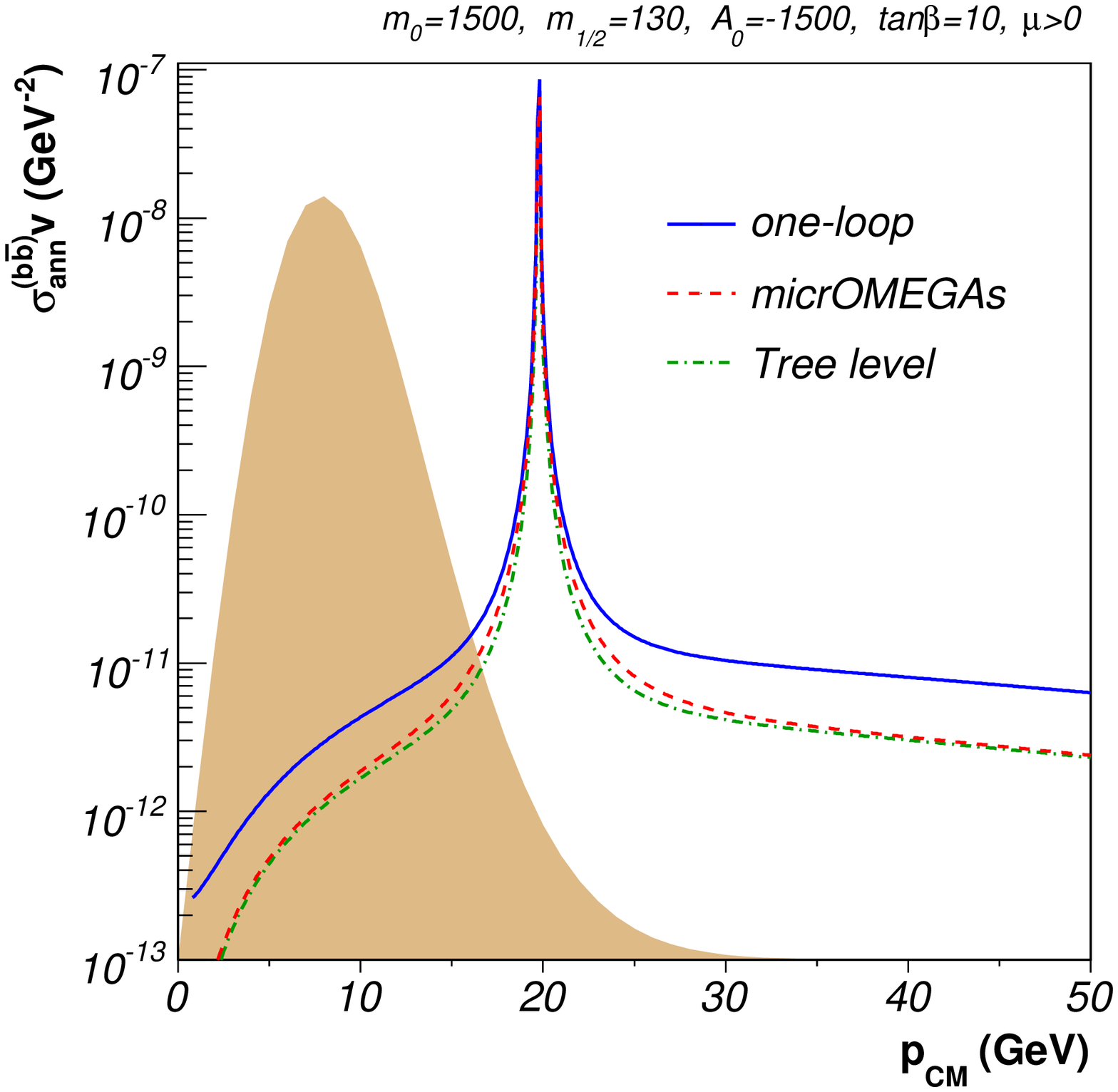}}}
		\put(155,152){\resizebox{!}{5cm}{\includegraphics{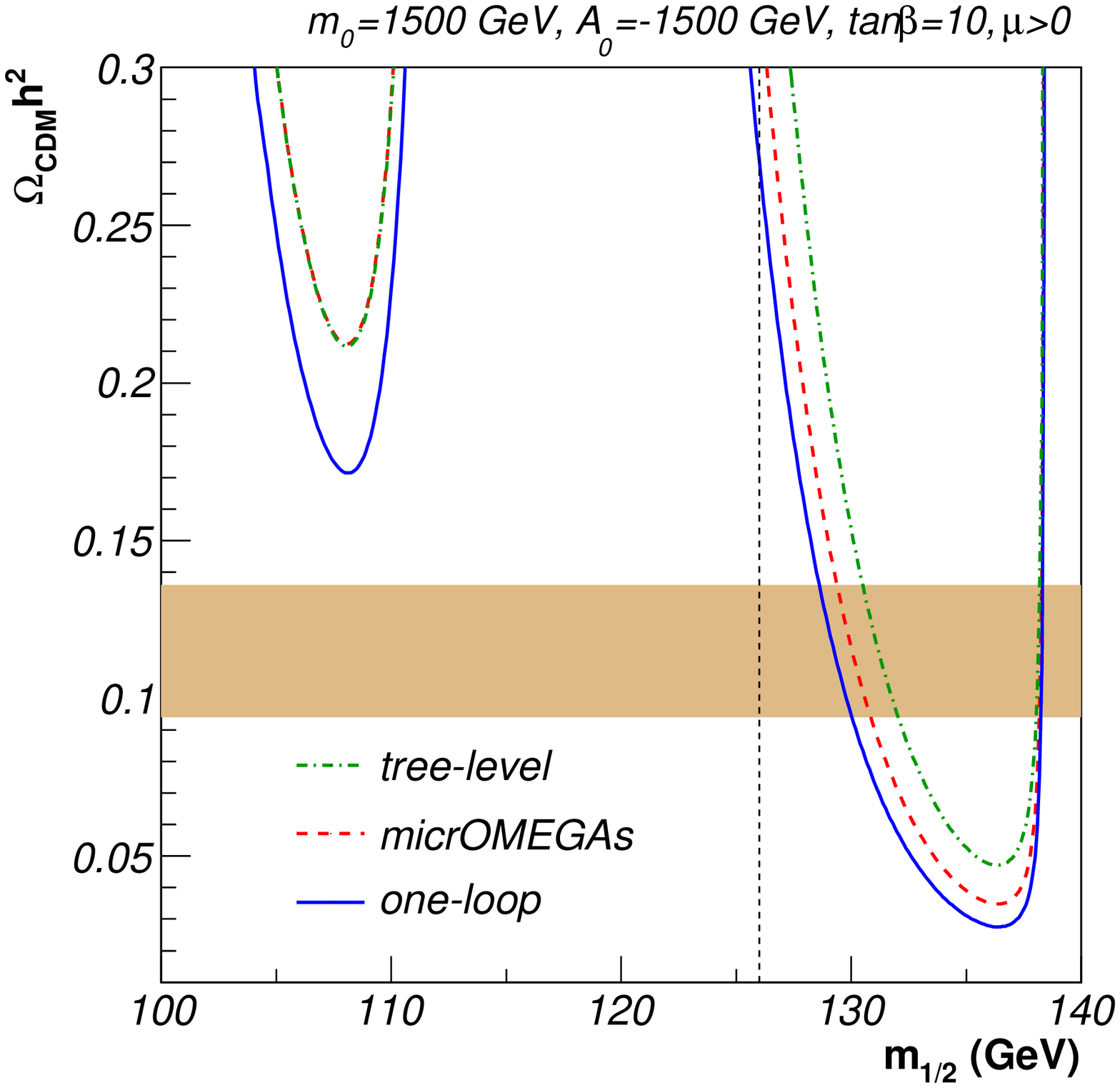}}}
		\put(5,0){\resizebox{!}{5.1cm}{\includegraphics{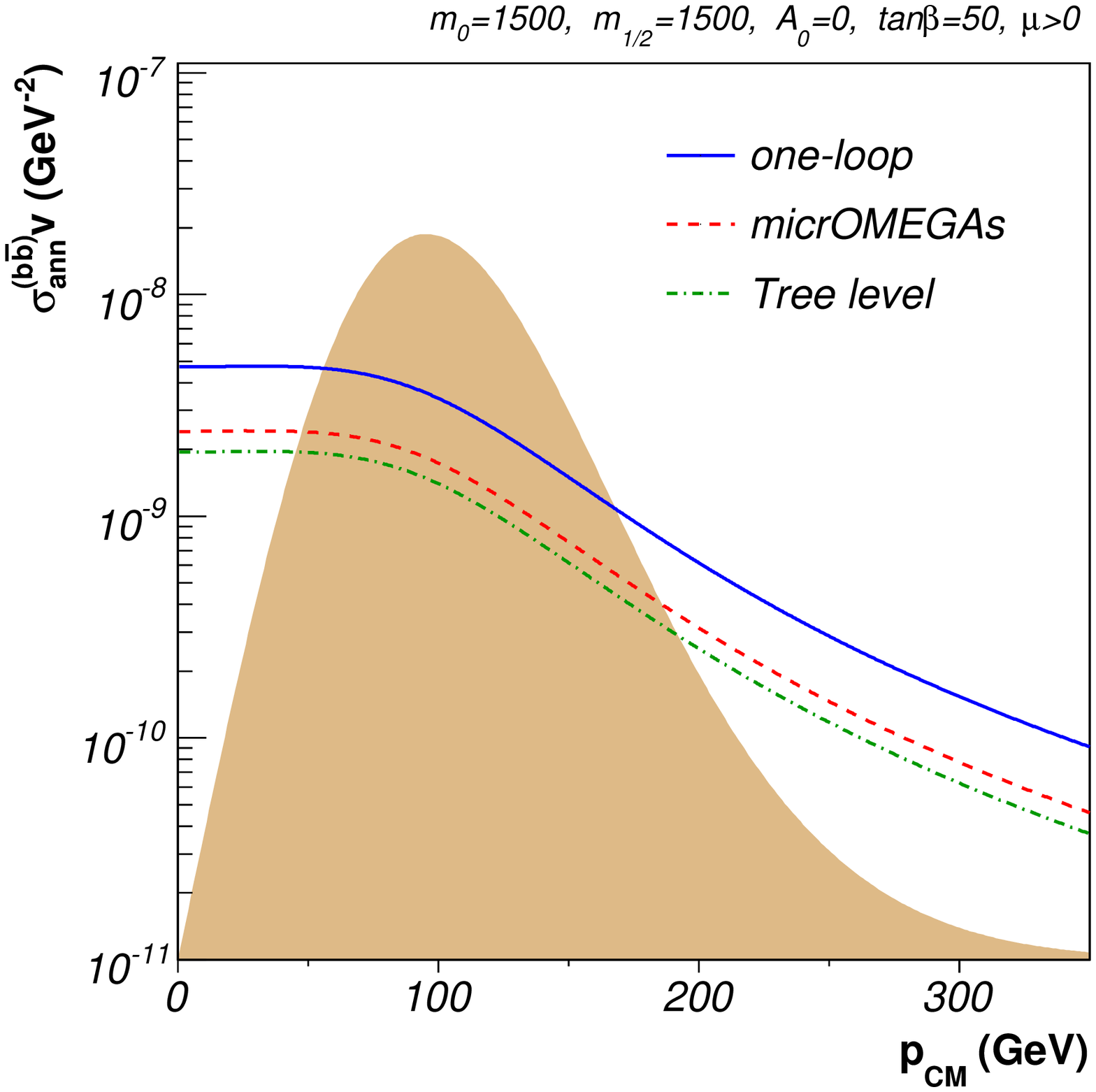}}}
		\put(155,2){\resizebox{!}{5cm}{\includegraphics{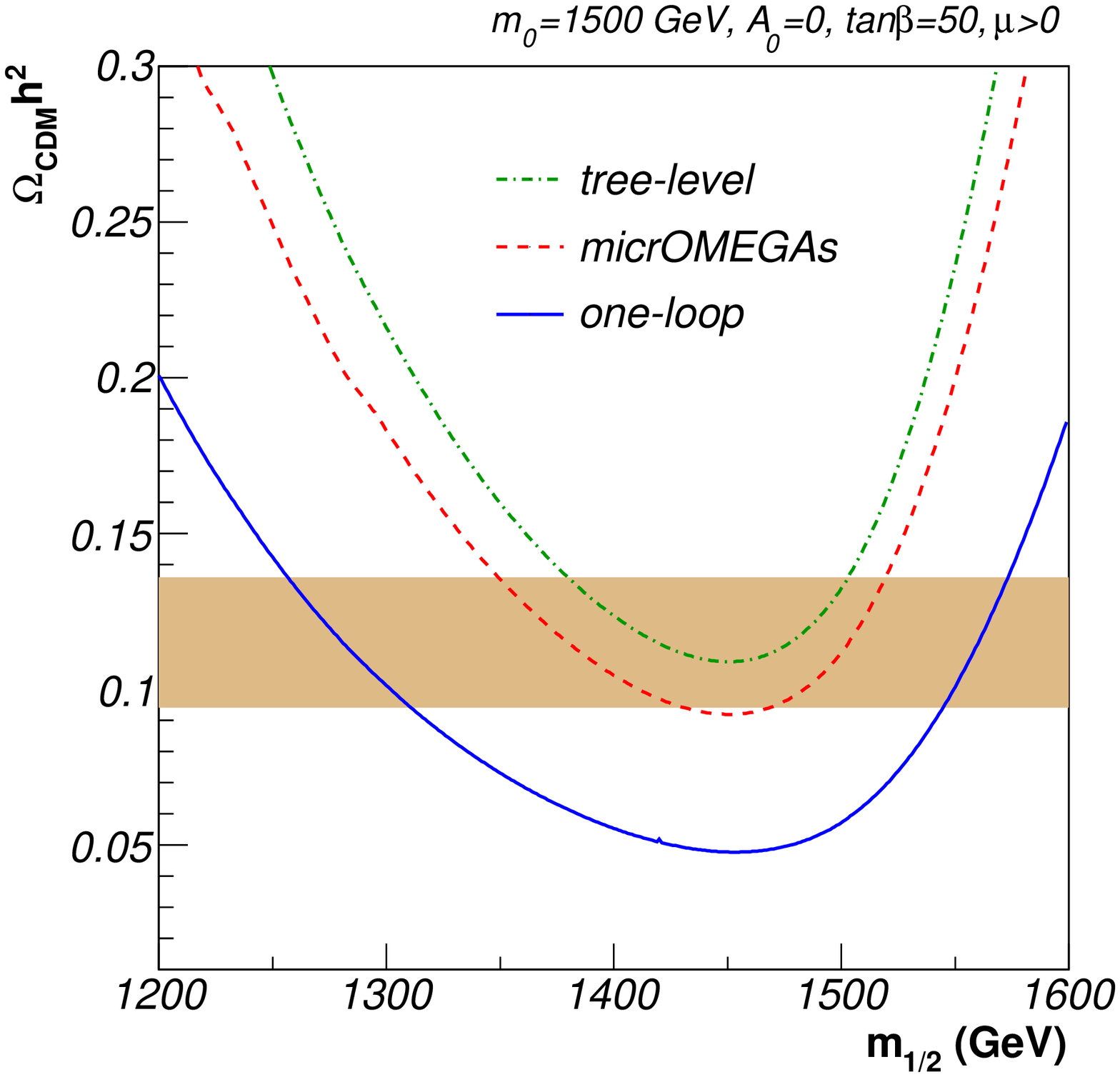}}}
		\put(310,2){\resizebox{!}{5cm}{\includegraphics{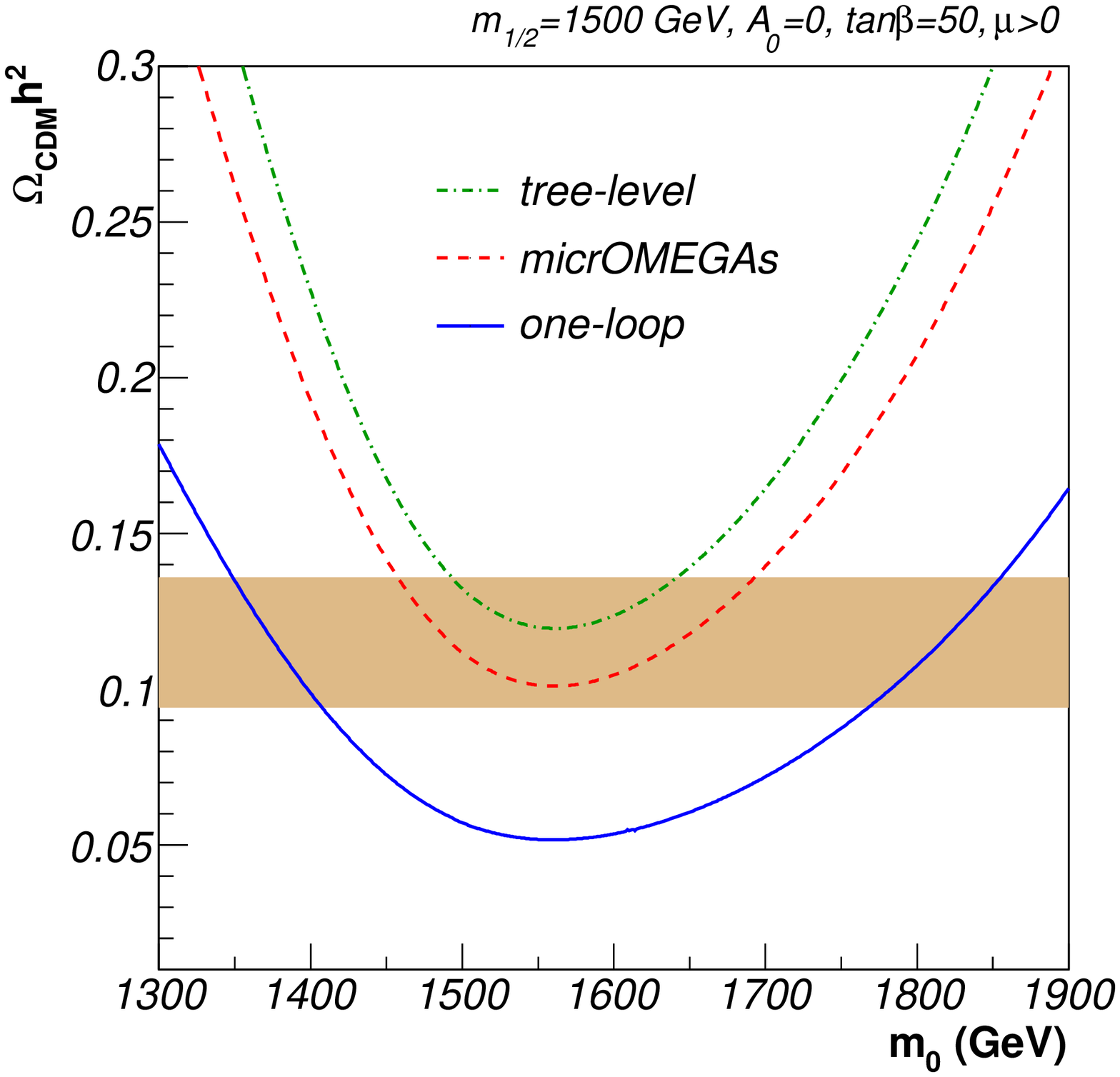}}}
	\end{picture}	
	\caption{The effects of radiative corrections on cross-section as a function of centre-of-mass momentum (left) and the effects on the resulting dark matter's relic density as a function of mSUGRA parameters $m_{1/2}$ (middle), $m_0$ (right). The results are shown for parameter points 1 and 2 with a dominant annihilation into bottom quarks.}
\label{Fig1}
\end{figure}
\section{RESULTS}
\begin{figure}[t]
	\begin{picture}(450,290)(0,0)
		\put(5,150){\resizebox{!}{5.1cm}{\includegraphics{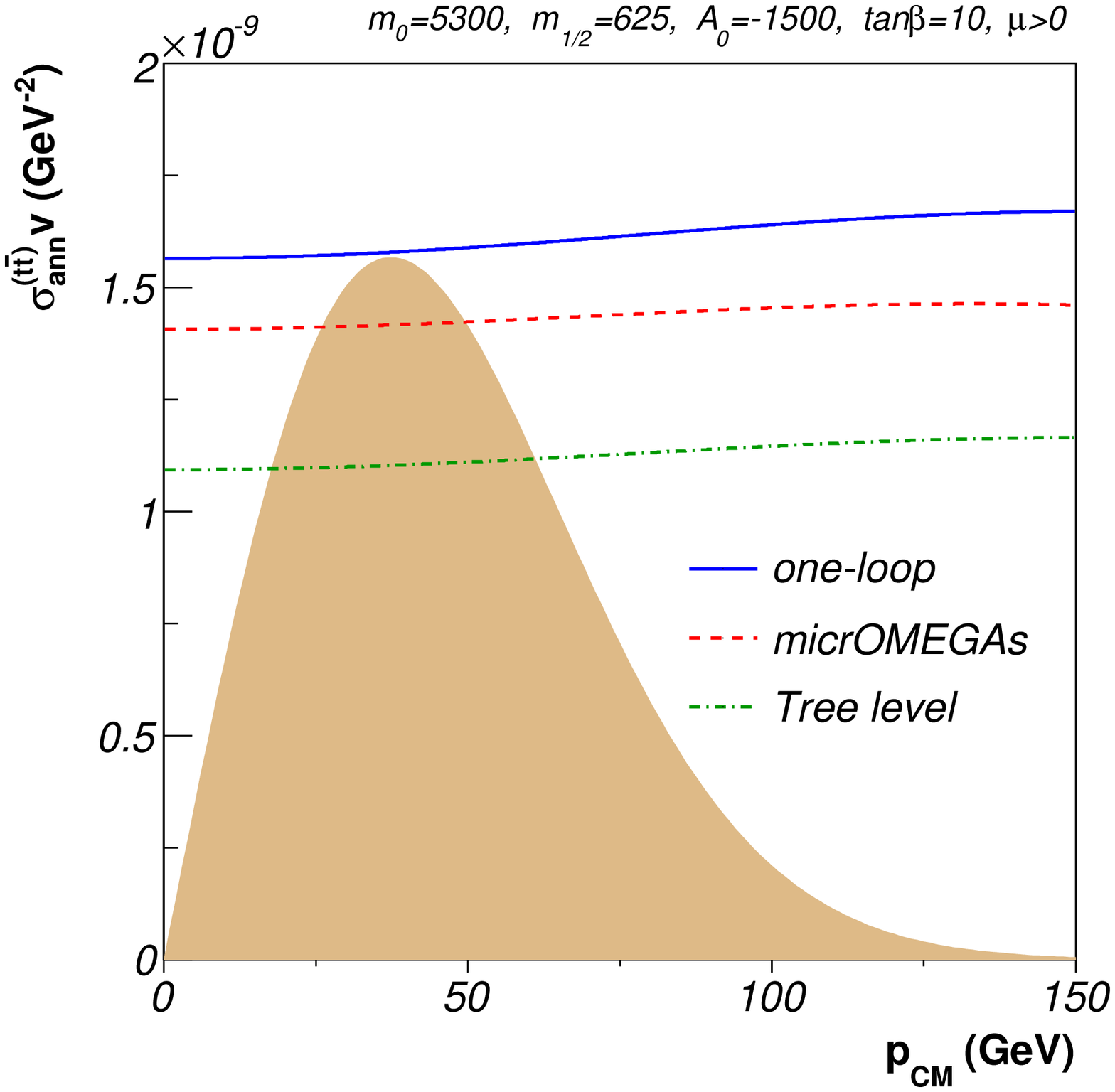}}}
		\put(155,152){\resizebox{!}{5cm}{\includegraphics{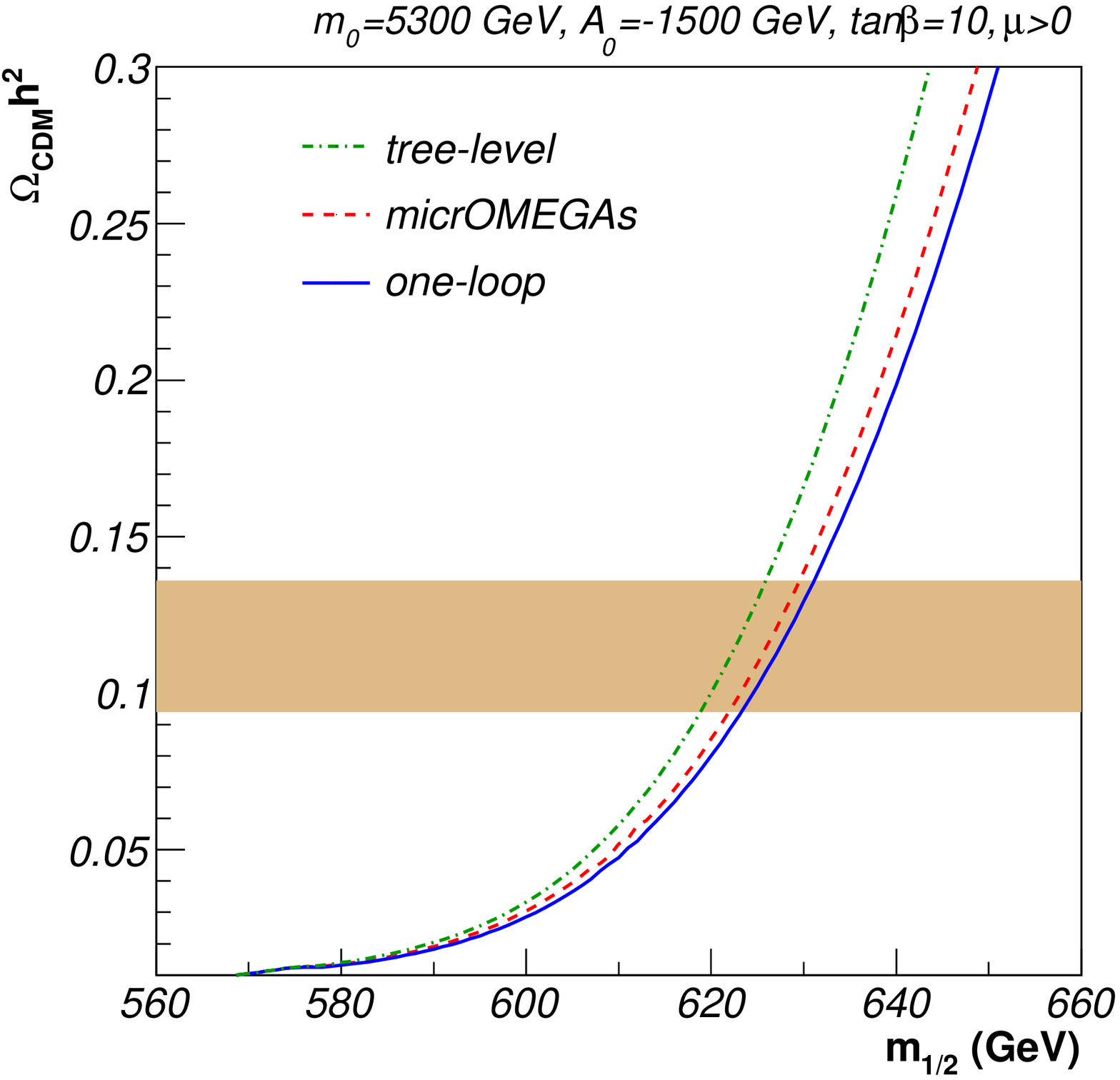}}}
		\put(310,152){\resizebox{!}{5cm}{\includegraphics{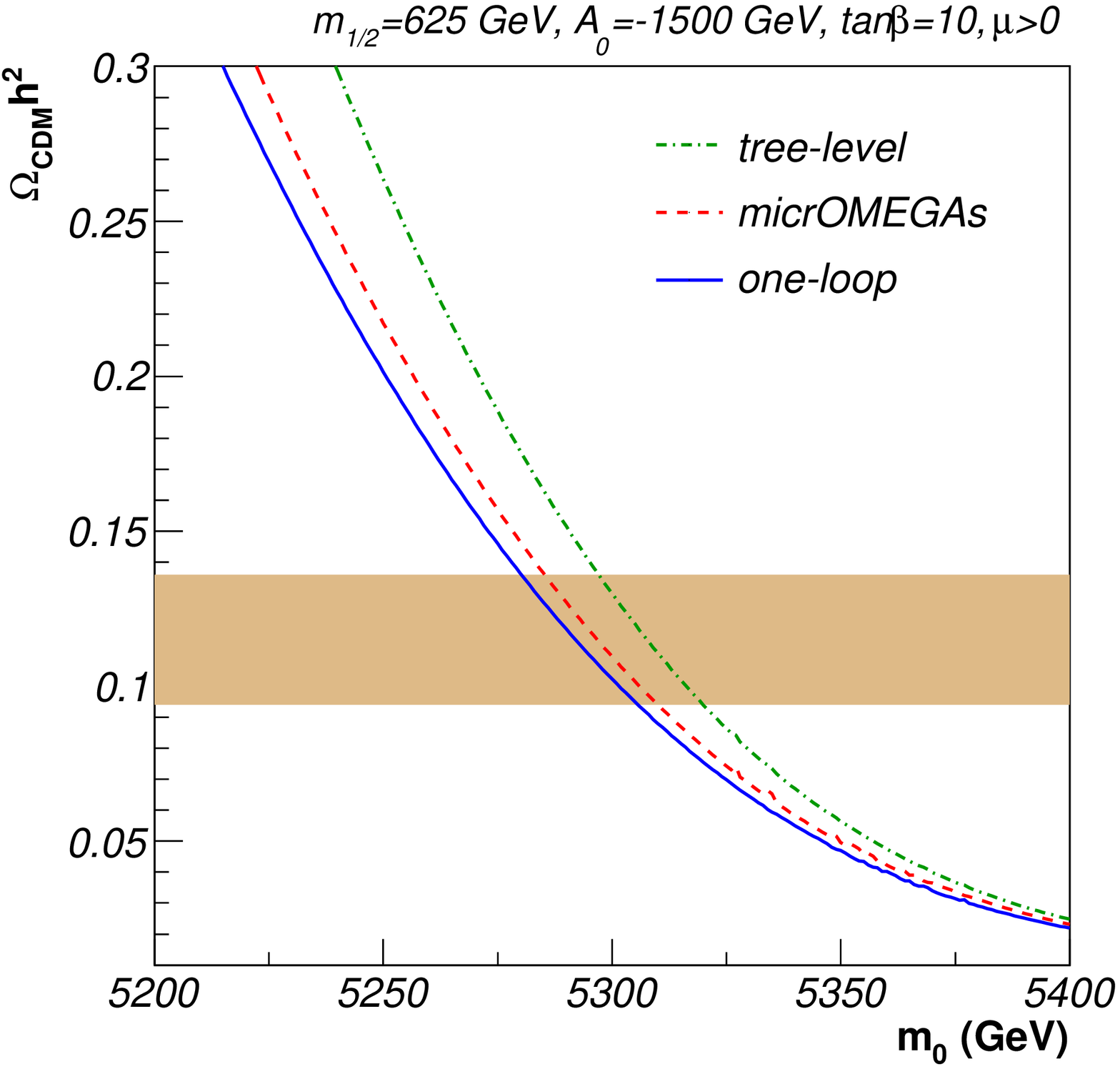}}}
		\put(5,0){\resizebox{!}{5.1cm}{\includegraphics{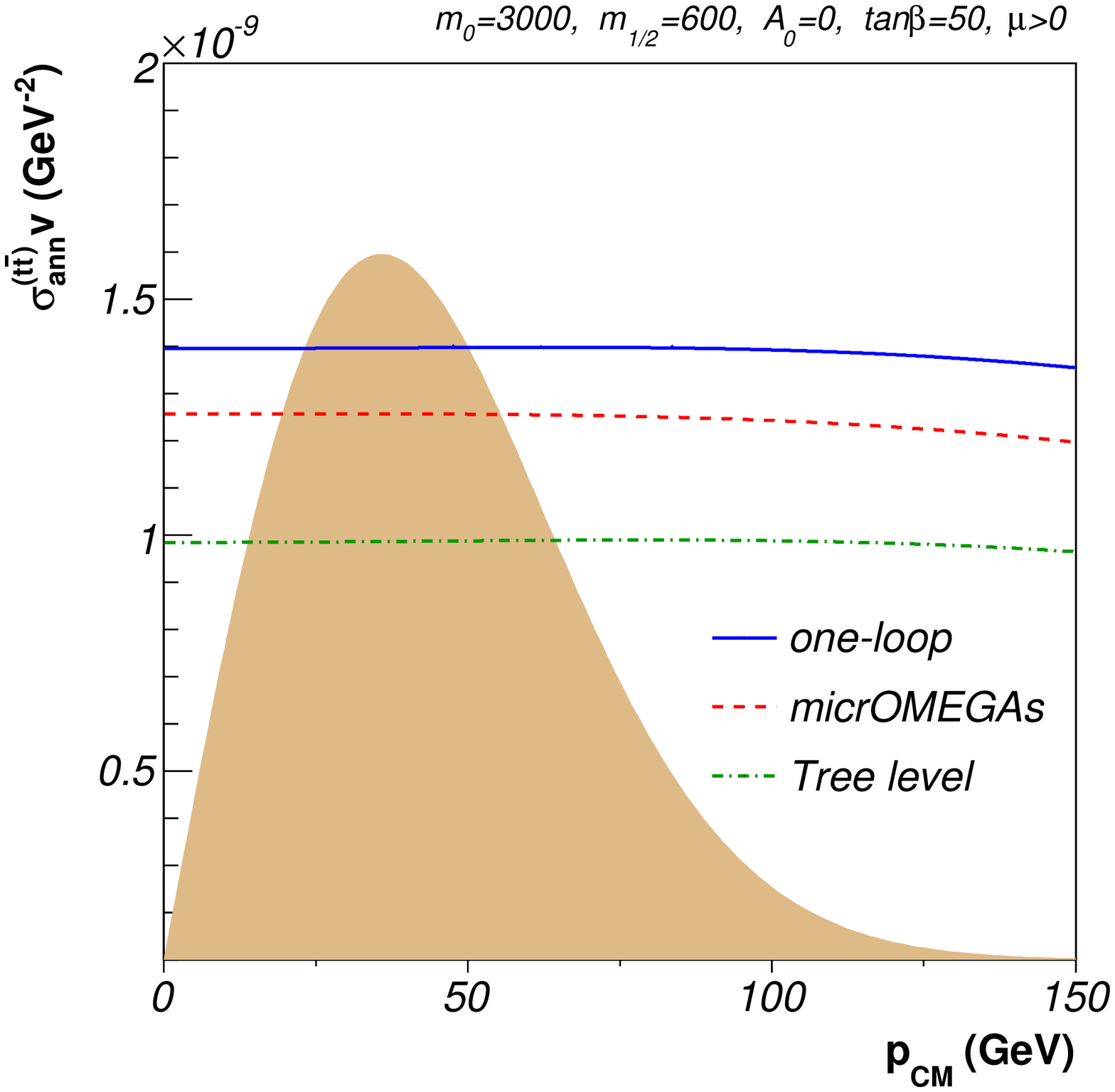}}}
		\put(155,2){\resizebox{!}{5cm}{\includegraphics{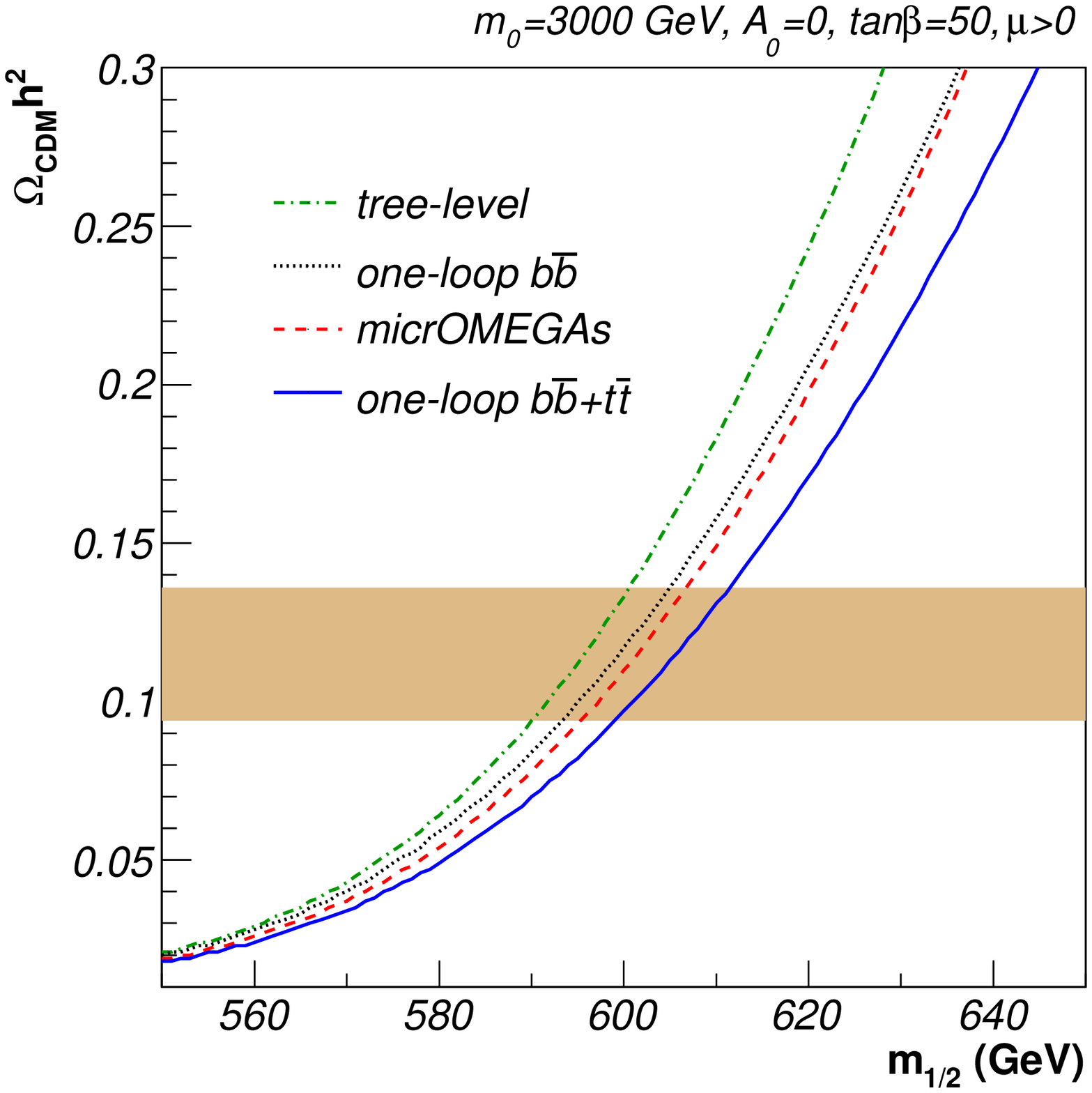}}}
		\put(310,2){\resizebox{!}{5cm}{\includegraphics{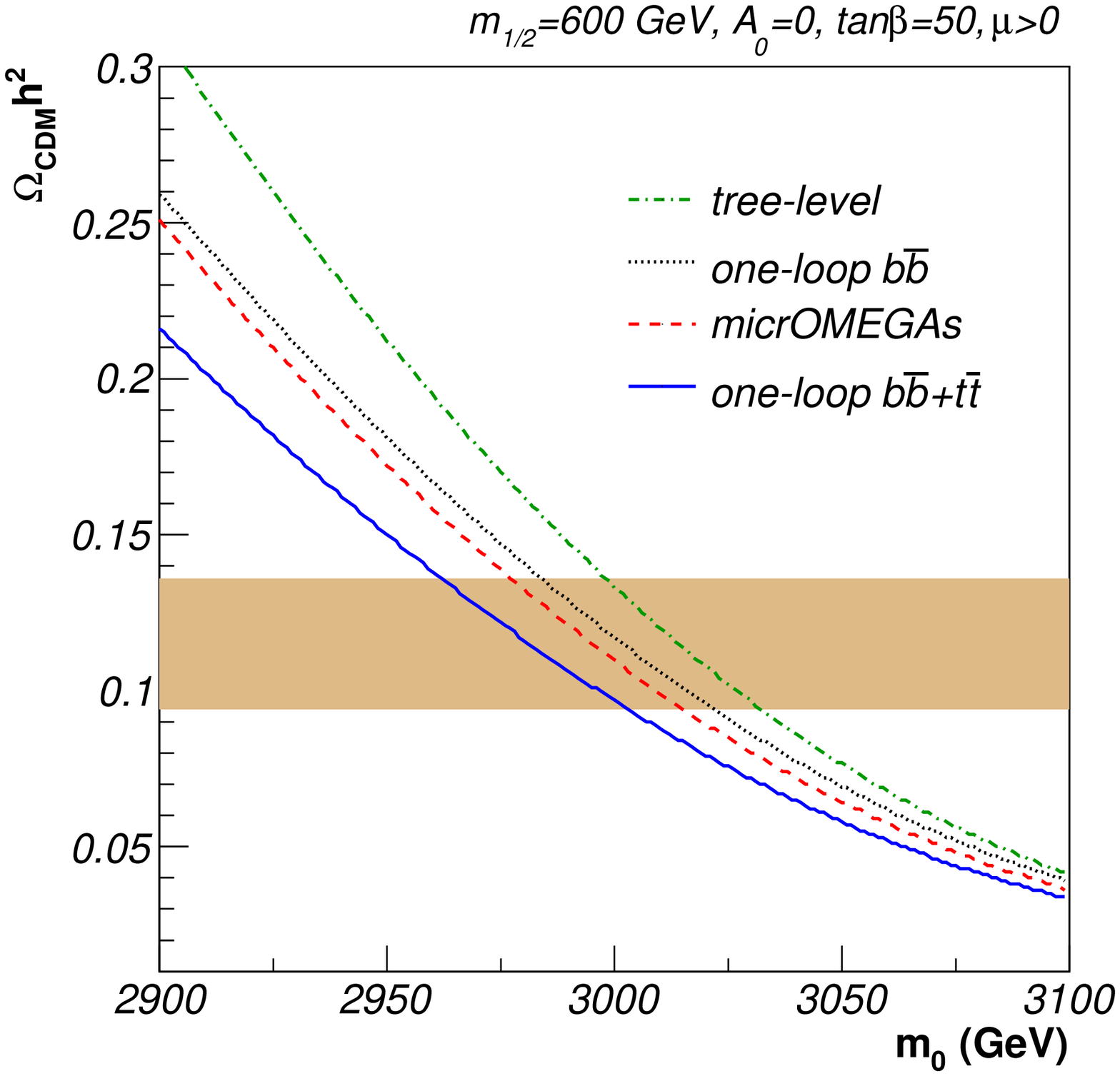}}}
	\end{picture}
	\caption{The effects of radiative corrections on cross-section as a function of centre-of-mass momentum (left) and the effects on the resulting dark matter's relic density as a function of mSUGRA parameters $m_{1/2}$ (middle), $m_0$ (right). The results are shown for parameter points 3 and 4 with a dominant annihilation into top quarks.}	
\label{Fig2}
\end{figure}

We calculate the complete QCD and SUSY-QCD corrections not only to the Higgs boson exchange but also to the $s$-channel Z-boson exchange and the $t$-channel and $u$-channel squark exchange. We also include parts of the correction for the Higgs exchange which are not treated in \cite{carena}. We find that the additional correction when including the full one-loop radiative result are 10-15\% relative to the corrections already included in {\tt MicrOMEGAs}.  
\newline %
We present numerical results for four typical mSUGRA parameter points shown in Tab.~\ref{Tpoints}. The high-scale parameters were evolved down to the electroweak scale using {\tt SPheno} \cite{spheno} and the neutralino relic density has been calculated with {\tt micrOMEGAs}, where we have included our calculation of the annihilation cross
section as discussed above. For each point, the relic density fulfills the limits (\ref{cWMAP}) and additionally, the branching ratio of the decay $b\to s\gamma$ lies within the experimental limits 
\begin{equation}
	{\rm BR}(b \to s \gamma) ~=~ \big( 3.55 \pm 0.26 \big) \cdot 10^{-4}\,,	
\end{equation}
obtained from combined measurements from BaBar, Belle, and CLEO \cite{bsgamma}. This limit leads to constraints on parameters connected to sfermions and gauginos.
\newline %
The selected points with a dominant bottom quark final state are either near the bulk region (point 1) with low fermion mass parameter $m_{1/2}$ for low $\tan\beta$ (the scalar mass is rather large in order to avoid the constraint by $b\to s \gamma$) or in the Higgs-funnel region for large $\tan\beta$ where our point 2 is positioned exactly on the CP-odd Higgs resonance. Both points 3 and 4 for which the top quark final state is important, lie in the focus point region where the scalar mass parameter $m_0$ is large. 
\newline %
In Figs.~\ref{Fig1} and \ref{Fig2} we show the numerical results for the
relevant annihilation cross section and the resulting neutralino relic
density for the points presented in Tab.~\ref{Tpoints}. We show the leading order
result (green dash-dotted), the approximation already implemented in
{\tt micrOMEGAs} (red dashed), and the result including our full
one-loop SUSY-QCD corrections (blue solid). In the graphs showing the
cross section (left row) we also indicate the Boltzmann distribution
function involved in the calculation of the thermal average (couloured
area).
\newline %
As already stated above, the one-loop contributions that are not
implemented in {\tt micrOMEGAs} lead to an increase of about 10 -- 15 \%
in the annihilation cross section both for bottom and top quark final
states. In consequence, the neutralino relic density decreases by about
the same amount for the different points, as can be seen in the center
and right panels showing the neutralino relic density as a function of
the high-scale mass parameters $m_{1/2}$ and $m_0$, respectively. For
fixed values of $\tan\beta$, $A_0$ and ${\rm sgn}\,\mu$, the favoured band in
the $m_0$-$m_{1/2}$ plane is shifted to lower gaugino masses and higher
scalar masses (focus point region).
\begin{acknowledgments}
The work of K.K. is supported by the ANR project ANR-06-JCJC-0038-01 and the work of B.H. was supported by a MENRT PhD. grant.
\end{acknowledgments}


\begin{thebibliography}{9}   
\bibitem{WMAP}
  J.~Hamann, S.~Hannestad, M.~S.~Sloth and Y.~Y.~Y.~Wong,
  Phys.\ Rev.\  D {\bf 75} (2007) 023522
  [arXiv:astro-ph/0611582].
%
\bibitem{Dsusy}
  P.~Gondolo, J.~Edsjo, P.~Ullio, L.~Bergstrom, M.~Schelke and E.~A.~Baltz,
  JCAP {\bf 0407} (2004) 008
  [arXiv:astro-ph/0406204],
%
\bibitem{microm}
  G.~Belanger, F.~Boudjema, A.~Pukhov and A.~Semenov,
  arXiv:0803.2360 [hep-ph],
G.~Belanger, F.~Boudjema, A.~Pukhov and A.~Semenov,
Comput.\ Phys.\ Commun.\  {\bf 176} (2007) 367
[arXiv:hep-ph/0607059],
G.~Belanger, F.~Boudjema, A.~Pukhov and A.~Semenov,
Comput.\ Phys.\ Commun.\  {\bf 174} (2006) 577
[arXiv:hep-ph/0405253],
G.~Belanger, F.~Boudjema, A.~Pukhov and A.~Semenov,
Comput.\ Phys.\ Commun.\  {\bf 149} (2002) 103
[arXiv:hep-ph/0112278].
%
\bibitem{drees}
  M.~Drees and M.~M.~Nojiri,
  Phys.\ Rev.\  D {\bf 47} (1993) 376
  [arXiv:hep-ph/9207234].
\bibitem{jung_rev}
  G.~Jungman, M.~Kamionkowski and K.~Griest,
  Phys.\ Rept.\  {\bf 267} (1996) 195
  [arXiv:hep-ph/9506380].
\bibitem{carena}
  M.~S.~Carena, D.~Garcia, U.~Nierste and C.~E.~M.~Wagner,
  Nucl.\ Phys.\  B {\bf 577} (2000) 88
  [arXiv:hep-ph/9912516].
\bibitem{drehik}
  M.~Drees and K.~i.~Hikasa,
  Phys.\ Lett.\  B {\bf 240} (1990) 455
  [Erratum-ibid.\  B {\bf 262} (1991) 497].
\bibitem{spheno}
  W.~Porod,
  Comput.\ Phys.\ Commun.\  {\bf 153} (2003) 275
  [arXiv:hep-ph/0301101].
\bibitem{bsgamma}
  E.~Barberio {\it et al.}  [Heavy Flavor Averaging Group (HFAG)],
  arXiv:hep-ex/0603003.


\end{thebibliography}
\end{document}